\documentclass[runningheads]{llncs}
\usepackage[T1]{fontenc}
\usepackage{graphicx,color}
\usepackage{setspace}

\begin{document}
\title{The Role of In-House Procurement According to Finnish Municipalities' Purchase Invoice Data}
\titlerunning{Public Procurement in Finnish Municipalities} 
%
\author{Reetta-Kaisa Ghezzi \inst{1} \and Minnamaria Korhonen\inst{2} \and
Hannu Vilpponen\inst{1} \and
Tommi Mikkonen\inst{1}}
\authorrunning{R. Ghezzi et al.}
%
\institute{University of Jyväskylä, Jyväskylä, Finland \and
Association of Finnish Municipalities, Helsinki, Finland\\
\email{reetta.k.ghezzi@jyu.fi, minnamaria.korhonen@gmail.com, hannu.v.vilpponen@jyu.fi, tommi.j.mikkonen@jyu.fi}}

\maketitle              
\begin{abstract}
Public sector is a large consumer for ICT systems and services development used for various public services. Tendering for such systems is governed by laws aimed at eliminating unfair advantages and offering all possible actors similar opportunities to participate in the tendering process. In this paper, we study in-house procurement, where the acquiring organization is an owner of the subcontractor that delivers the system. Municipalities’ purchase in-voice data is used to determine how greatly municipalities in Finland depend on in-house procurement. In conclusion, the understanding if included municipalities have ICT service and development units within the organizations needs closer examination, as in-house companies may offer municipalities with limited resources divided costs in the public procurement process.

\end{abstract}
\section{Introduction}

Public procurement of Information and Communication Technology (ICT) systems and services is large, and international companies and governments at national, regional, state and city levels are major buyers of ICT goods and services \cite{arrowsmith2002public}. As an example, the state of Finland alone made procurement worth over EUR 1000 million \cite{tivi-10M} on ICT during the year 2020. 

The public procurement process is mandated by the EU and national procurement legislation within the EU \cite{directive-procurement}. Public ICT system development projects are often triggered by EU or national legislation, which sets a time frame for the development, which must be followed under the threat of sanctions \cite{vilpponen_2021}. The preparation of the tender is a heavy and complex process that requires 
resources and the necessary expertise at the very beginning of procurement planning stage \cite{vilpponen_2021}. 
Furthermore, strict parameters set by laws and directives hinder the effectiveness of the procurement \cite{keranen-procurement}. According to the Finnish Procurement Act \cite{FINprocurement}, the procurement can be taken to the Market Court for resolution, which typically delays the procurement process by up to a year.

An assumption is, that in many cases, the lack of resources needed for the tendering process can be solved with in-house procurement. In-house companies are owned and controlled by the state, municipality or a group of municipalities, from which procurement can be carried out without tendering in accordance with the Procurement Act \cite{iloranta-hankintojenjohtaminen}. With this setup, the public administration can avoid lengthy and uncertain procurement processes, to stay within the deadlines set by the legislation. Insufficient procurement resources may also affect the desire to take advantage of the opportunity to use in-house companies to procure ICT goods or services.

In this paper, we examine in-house purchase data to determine how greatly municipalities in Finland depend on in-house procurement. Furthermore, we examine what could be the adequate ways to explore the reasons, why municipalities use or do not use in-house procurement. 

\section{Background and Motivation}

Finland is a Nordic EU member state with 5,5 million citizens. Hartung and Kuźma \cite{hartung_kuźma_2018} depict that Procurement Directive within European Union gives opportunities for the member states to determine how to implement in-house procurement into the legislation. At least Finland and Poland have in-house procurement implemented in procurement legislation, whereas other member states have not included it in their legislation \cite{hartung_kuźma_2018}. 

Public agencies have an option to buy in-house, which is not in the circle of the Public Procurement Act \cite{FINprocurement}. In-house procurement is an interesting way to buy ICT goods and services because the owner of the in-house company does not need to follow procurement procedures in acquisition, which is a significant and allowed derogation to the Public Procurement Act \cite{FINprocurement}. An in-house company must be owned by different procurement units, and it can only have limited business with parties other than the owners from outside sources \cite{FINprocurement}. At present, an in-house unit may have a five percent or a maximum of EUR 500 000 business outside the procurement units \cite{FINprocurement}. However, the limit for outside business is ten percent, and the EUR 500 000 restriction evaporates if the market cannot provide the needed service  \cite{FINprocurement}. Furthermore, the in-house position also requires that the owner agency has a deputy in the in-house company to ensure the decision-making authority \cite{FINprocurement}. 

The phenomenon is interesting in Finland where the number of municipalities is as high as 309, and public sector owns over 2000 in-house companies and Finnish Competition and Consumer Authority (FCCA) estimates that the overall turnover in Finnish in-house companies in all sectors is yearly over EUR 40 billion \cite{suomen_yrittajat_2022}. The public discussion in Finland has focused on changing the Procurement Act to even more allowing direction in in-house procurement by raising the limits to sell goods and services outside the in-house \cite{suomen_yrittajat_2022}. This has inflicted a counter reaction in the regular market. Currently, five percent limit to sell outside the in-house means, that for example from EUR 15 million turnover in-house company can make sales worth EUR 750 000 outside the stakeholders. If the limit was to be raised in 20\%, which has been suggested in Finland, the sales outside the in-house company's stakeholders would raise substantially. The current threshold for EUR 500 000 is suggested to be removed. With these estimates, the sales outside all the in-house companies in Finland may grow up to eight billion euros. This is problem-some because the municipalities and other stakeholders can purchase goods and services from the in-house company without following Public Procurement Act. 
%



The Procurement Act only applies to procurement that exceeds EU thresholds and national thresholds \cite{FINprocurement}. The Procurement Act does not apply to so-called small procurement that fall below the threshold values. In Finland, the national threshold value is EUR 60 000 for estimated costs over four years \cite{FINprocurement}. Typically, even medium-sized ICT system procurement exceeds the national or EU threshold value, in which case tendering has to be done. 

One key form of operation of in-house companies is organizing public tenders, where they tender ICT products and services from commercial vendors. Since the in-house companies have a large owner base, it is possible to get favorable offers based on the tender volume. The public administration can acquire the purchased ICT systems and services without tendering by using its ownership in the in-house company. The aim is to determine the scale for the in-house purchases, which can be evaluated through public agency purchase invoices.

\begin{figure}[!t]
    \centering
    \includegraphics[width=0.85\textwidth]{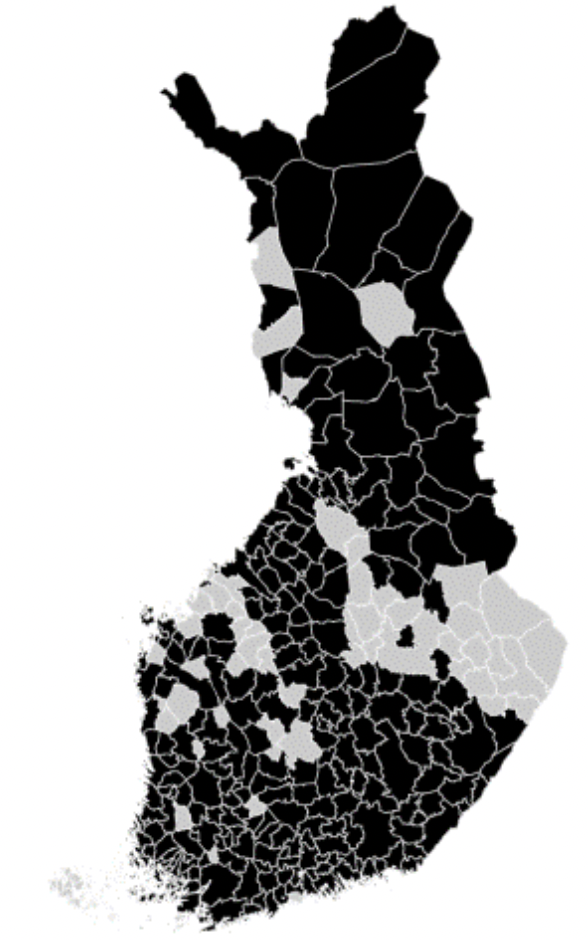}
   \caption{Kuntien Tiera, depicted geographically with black color has 403 owners including municipalities, joint authorities and  other in-house companies \cite{lisenssi}. The owners are listed in Table \ref{tab:in-house}.}
    \label{fig:owners}
\end{figure}

As an example, the left-hand side of Figure \ref{fig:owners} presents the geographical coverage for one in-house, Kuntien Tiera, in Finland. The right-hand side of the figure presents some key data regarding the studied in-house organizations.
    

\begin{table}[ht]
\begin{center}
\caption{All the studied organizations are characterized in the table below (T/O refers to annual turnover in million euros).} \label{tab:in-house}
\begin{tabular}{c|cccccc}
\textbf{In-house} & \textbf{T/O} & \textbf{Empl.} & \textbf{Owners} &  \\
\hline
Sarastia        & 15,4              & 914                & 284             &  \\
Istekki         & 15,3              & 833                & 63              &  \\
2M IT             & 11,8              & 515                & 47              &  \\
Monetra           & 77,6              & 1173               & \textgreater 42 &  \\
Kuntien Tiera  & 47,2               & 274                & 403             & 
\end{tabular}
\end{center}
\end{table}


\section{Research Approach}

In this paper, we examine in-house purchase invoice data to determine how greatly municipalities in Finland depend on in-house procurement. Furthermore, we examine what could be the adequate ways to explore the causes, why municipalities use or do not use in-house procurement. 

The focus groups in this research are the largest municipalities and largest ICT in-house companies in Finland. Opening the purchase invoice data is voluntary. Eleven largest municipalities in Finland have opened the purchase invoice data, hence, these receive closer examination. 

Municipalities' purchase invoice data is available Avoindata.fi \cite{avoindata_2022} and at the municipalities' websites. In addition, invoice purchase data inquiries were made directly to the municipalities when the information was ambiguous or non-complete. Shareholder information was searched on the websites of in-house companies and inquired by e-mail, and in some cases additionally by the phone. Public documents such as shareholder agreements were used to determine whether the in-house position existed. 
Another limitation involves the purchase invoice accounts. AURA is an official handbook maintained by Ministry of Finance that gives guidelines for mandatory economic reporting of municipalities and joint municipal authorities \cite{aura_2022}. It contains a mandatory level of chart of accounts, and it was used to ensure comparability between the accounts of different municipalities. All the ICT related purchases were chosen to be inspected. 
The full calendar year 2021 is chosen to be the reference year for all municipalities. 


\section{Results}

A summary of 11 largest municipalities in Finland is presented in Table \ref{tab:my-table}, together with five largest in-house companies and their monetary value per municipality included. 
Based on the findings, municipalities have varying ways to acquire ICT solutions. Whether the larger municipalities rely more on traditional ICT procurement from the market, whereas the smaller cities use eagerly in-house procurement, remains unresolved and demands more research. Overall, limited resources altogether might lead towards in-house procurement. 
The more precise analysis of the data is presented through Primary Empirical Conclusions (PECs), based on purchase invoice data and other necessary documents, presented above. The PECs we have formulated are the following:

\begin{table}[t]
\begin{center}
\caption{Purchases from chosen in-house companies by 11 biggest municipalities in Finland, based on open purchase invoice data year 2021. *VAT included; **Not owner.}
\begin{tabular}{c|cccccc}
Municipality & Istekki & Monetra & Kuntien Tiera & Sarastia  & 2M-it & Total 1000 € \\
\hline
Helsinki     & -          & 1            & 2                & 240          & -        & 243          \\
Espoo        & -           & -            & 1 321            & 1 282        & 144      & 2 747        \\
Tampere      & -          & -            & -                & -            & -        & -            \\
Vantaa       & -          & \textless{}1 & 798              & 104          & -        & 902          \\
Oulu         & 197        & 5 471        & 195               & -            & -        & 5 862        \\
Turku       & -          & 1            & 11 901           & 5 695        & 4 935    & 22 532       \\
Jyväskylä    & -          & 6 414        & 549              & -            & -        & 6 963        \\
Kuopio*       & 21 820     & 4 950       & -                & -            & -        & 26 770       \\
Lahti        & 3**         & 1            & 216              & \textless{}1** & -        & 219          \\
Pori         & -          & 1            & 417              & 3 174        & 5 993    & 9 584        \\
Kouvola     & -          & -            & 67               & 2 414        & -        & 2 481        \\
Total €      & 22 019     & 16 837       & 15 465           & 12 909       & 11 072    & 78 303     
\end{tabular}
\end{center}
\label{tab:my-table}
\end{table}
 
\textbf{PEC1: Municipal autonomy.} 
Municipalities in Finland have autonomy to acquire independently solutions from the in-house companies or from the free market which is visible in the results. 
Table \ref{tab:my-table} presents the differences in relying on in-house procurement. Municipal autonomy results in non-standardized ICT purchasing practices, and as imaginable, the variety in solutions is high, which may lead to challenges in building effective and inter-operable systems.
Moreover, based on the results, it seems like some municipalities have direct access to the best products in the market, whereas some others need to adopt what suits the best for the group of stakeholders that own the in-house company. Understanding if included municipalities have ICT service and development units within the organizations needs closer examination, as in-house companies may offer municipalities with limited resources divided costs in the public procurement process. 

%


\textbf{PEC2: Selection of the supplier.}
The chosen accounts reveal that each municipality purchases accounting services in centralized manner from one of their own in-house companies. 
The result is sensible. Repetitive task such as accounting services are more manageable if one company provides them. 

\textbf{PEC3: In-house position through ownership and decision-making power.} In-house companies own shares of other in-house companies. The explanation lies in the shareholder agreement. For example, Monetra's shareholder agreement reveals that the in-house position is ensured to some extend through cross-ownership between the in-house companies. However, the cross-ownership might not be enough to ensure in-house position, and the decision-making power is necessary at least as a legislative demand \cite{FINprocurement}.



\textbf{PEC4: Un-used ownership.} Some owners have no purchases from their  in-house companies. For example, 
Tampere holds in-house position and decision-making power in Monetra, but has no related purchases. 
The ownership may have originated from previous contracts, acquisitions and mergers.  

\textbf{PEC5: Difficulties in gathering the data.} Half of the in-house companies provide complete shareholder information. Generally, the in-house companies responded to the inquiries with interest, and provided the names of the owners. Sarastia, Istekki and Meita provided full shareholder information including the names and division of the shares. In these cases, the relationships between the municipalities and in-house companies are rather easy to determine. 

\section{Conclusions}

Examining the relationship between Finnish municipalities and in-house companies reveals several future research areas. 
%
In-house procurement is more common for smaller municipalities. It seems, that many of the municipalities buy shares from the in-house companies to prepare themselves for the future needs resource-wise. In addition, larger municipalities may have wider ICT services and development units within the organization. 

In this article, the in-house position was sometimes demanding to determine. It seems that determining the in-house position is demanding for the municipalities as well. Finnish Competition and Consumer Authority (FCCA) has made a decision on inadequate interpretation on in-house position \cite{kkv_kauppakamari}. Municipality or its stakeholders need to participate in strategic decision-making in the in-house company. If decision-making power is not adequate in the in-house company, the legal ramifications may follow. The criteria how municipalities and in-house companies interpret the in-house position contradicts to some extend with the FCCA interpretation. Interpretations on in-house position need more research. 
Finally, difficulties in gathering the data and receiving information from in-house companies contradicts with the openness and equality principles. Half of the times, the information on shareholders in the in-house companies was not available. 

%
%
%
\bibliographystyle{unsrt}
%
\bibliography{references}

\end{document}